\documentclass[aps,twocolumn,groupedaddress,floatfix]{revtex4-1}
\usepackage{amssymb,amsmath}

\usepackage{graphicx}
\usepackage{subfigure}
\usepackage[english]{babel}
\usepackage{float}
\usepackage{color}
\usepackage[document]{ragged2e}

\usepackage{lipsum}
\usepackage{suffix}
\usepackage{mathtools}
\DeclarePairedDelimiterX\MeijerM[3]{\lparen}{\rparen}%
{\begin{smallmatrix}#1 \\ #2\end{smallmatrix}\delimsize\vert\,#3}

\newcommand\MeijerG[8][]{%
  G^{\,#2,#3}_{#4,#5}\MeijerM[#1]{#6}{#7}{#8}}

\WithSuffix\newcommand\MeijerG*[7]{%
  G^{\,#1,#2}_{#3,#4}\MeijerM*{#5}{#6}{#7}}
\begin{document}
\newcommand{\be}{\begin{equation}}
\newcommand{\ee}{\end{equation}}
\newcommand{\rojo}[1]{\textcolor{red}{#1}}

\title{Fractionality and $\cal{PT}$- symmetry in an electrical transmission line}

\author{Mario I. Molina}
\affiliation{Departamento de F\'{\i}sica, Facultad de Ciencias, Universidad de Chile, Casilla 653, Santiago, Chile}

\date{\today }

\begin{abstract} 
We examine the stability of a 1D electrical transmission line in the simultaneous presence of  ${\cal PT}$-symmetry and fractionality. The array contains a binary gain/loss distribution $\gamma_{n}$ and
a fractional Laplacian characterized by a fractional exponent $\alpha$. For an infinite periodic chain the spectrum is computed in closed form, and its imaginary sector is examined to determine the stable/unstable regions as a function of the gain/loss strength and fractional exponent. In contrast to the non-fractional case where all eigenvalues are complex for any gain/loss, here  we observe that a stable region can exist when gain/loss is small and the fractional exponent is below a critical value,  $0<\alpha<\alpha_{c_{1}}$. As the fractional exponent is decreased further, the spectrum acquires a gap with two nearly-flat bands. We also examined numerically the case of a finite chain of size $N$. Contrary to what happens in the infinite chain, here the stable region lies always above a critical value $\alpha_{c_{2}}<\alpha<1$. An increase in gain/loss or $N$ always reduces the width of this stable region until it disappears completely.
\end{abstract}

\maketitle

{\em Introduction}.  
Two recent physics developments have aroused increased attention in recent years. One, is the notion of ${\cal PT}$ symmetry and the other is the rebirth of the old idea of fractionality. ${\cal PT}$-symmetric systems are characterized for having a non-Hermitian Hamiltonian but a real spectrum. This is a nontrivial departure of the standard idea in quantum mechanics where the Hamiltonian is always hermitian in order to ensure real eigenvalues. A closer look to this concept revealed that, in order to possess real eigenvalues, the Hamiltonian only needs to be invariant under the combined operations of parity and time reversal\cite{bender1,bender2}. In quantum mechanics this translates into the requirement that the real part of the potential be an even function in space $V_{R}(-x) = V_{R}(x)$, while its imaginary part be an odd function in space $V_{I}(-x)=-V_{I}(x)$. Typically, what happens is that as we increase $V_{I}(x)$, the spectrum remains real until a critical value of $V_{I}(x)$ is reached where two complex eigenvalues appear, spoiling the careful gain/loss balance of the system, whose dynamics becomes now unstable. This is known as a spontaneous symmetry breaking\cite{optics1}. 

Currently, numerous $\cal{P}\cal{T}$-symmetric systems have been explored in several contexts, from , solid state and atomic physics\cite{solid1,solid2}, optics\cite{optics1,optics2,optics3,optics4,optics5}, electronic circuits\cite{circuits}, magnetic metamaterials\cite{magnetics1,magnetics2},  and electrical transmission lines\cite{lazo},  among others. The ${\cal P}{\cal T}$ symmetry-breaking phenomenon has been observed in several experiments\cite{optics5,experiment2,experiment3}.

On the other hand, fractionality has experienced a rebirth from a mathematical curiosity, to a full-fledged research field. Roughly speaking, it consists of the idea of extending the notion of an integer derivative, to a fractional one. It dates back to the observation that the derivative $d^{n} x^{k}/d x^{n} = k!/(k-n)!\ x^{k-n}$ for integer $n$ could be extended to non-integer orders by means of the Gamma function: $d^{\alpha} x^{k}/d x^{\alpha} = \Gamma(k+1)/\Gamma(k-\alpha+1)\ x^{k-\alpha}$. From heuristic arguments such as this one, a whole field began to take shape thanks to the work of several noted mathematicians: Riemann, Euler, Laplace, Caputo, to name a few. To date there exists a number of well-defined definitions for fractional derivatives. One of the most used is the Riemann-Liouville form
\be
\left( {d^{\alpha}\over{dx^{\alpha}}} \right) f(x) = {1\over{\Gamma(1-\alpha)}} {d\over{dx}} \int_{0}^{x} {f(s)\over{(x-s)^\alpha}}\ ds,
\ee
where $0<\alpha<1$.  The non-local character of the fractional derivative has proven useful in a variety of fields: fluid mechanics\cite{quantum}, fractional kinetics and anomalous diffusion\cite{kinetics1,kinetics2,kinetics3}, strange kinetics\cite{strange}, fractional quantum mechanics\cite{frac1,frac2}, Levy processes in quantum mechanics\cite{levy}, plasmas\cite{plasmas}, electrical propagation in cardiac tissue\cite{cardiac}, cancer growth and invasions\cite{invasions}, and epidemics\cite{epidemics}, among others.
\begin{figure*}
\includegraphics[scale=0.21]{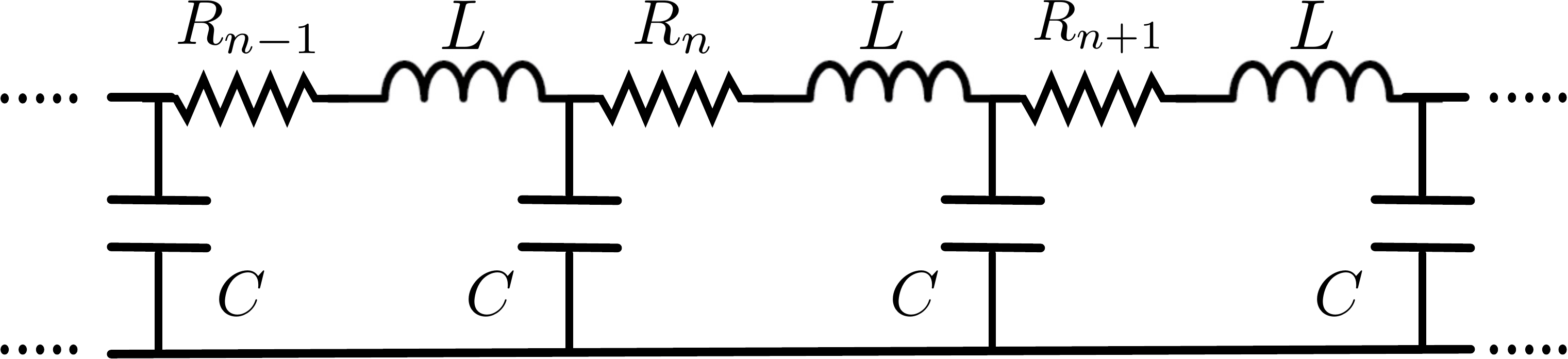}
\caption{A ${\cal PT}$-symmetrical electrical transmission line, with resistances distribution $R_{n}= (-1)^n R$.  }
\label{fig1}
\end{figure*}

In this work we examine the mutual interplay between ${\cal P}{\cal T}$ and fractionality in a 1D electrical  transmission line, trying to ascertain the stability properties as a function of `gain/loss' and fractionality. In a previous work\cite{molina} and using a standard tight-binding 1D chain, we found that in general, the presence of fractionality and ${\cal PT}$ tends to drive the system into instability. It is interesting then, to see what happens in a classical discrete periodic system, such as an electrical lattice. In particular, we seek to determine how the presence of fractionality with its nonlocal effects, affects the delicate gain and loss balance in a system that obeys the ${\cal P}{\cal T}$ condition. As we will see, in this system the presence of fractionality has a stabilizing effect on the system's spectrum. As the fractional exponent drops below a certain critical value, away from the standard, non-fractional value, a stability window opens in 
gain/loss - fractionality parameter space.

{\em The model.}\ \ We start from a 1D electrical transmission line, with resistance distribution given by $R_{n}=(-1)^n R$ (Fig.\ref{fig1}). Let us first assume no fractionality. Then, after use of Kirchoff's rules, we find the equation for the charges in each capacitor:
\be
{d^2\over{d t^2}} Q_{n}(t) + {R_{n}\over{L}} {d\over{d t}} Q_{n}(t) - {1\over{L C}} (Q_{n+1}-2 Q_{n} + Q_{n-1})=0
\ee
A stationary solution is sought in the form $Q_{n}(t)=Q_{n}\ \exp(i \Omega t)$:
\be
-\Omega^2 Q_{n}+ i \gamma_{n} \Omega Q_{n}-\omega^2(Q_{n+1}-2 Q_{n}+Q_{n-1})=0 
\ee
This equation can be split into two:
\begin{eqnarray}
-\Omega^2 a_{n} + i \gamma \Omega a_{n} - \omega^2(b_{n+1}-2 a_{n}+b_{n-1})&=&0\nonumber\\
-\Omega^2 b_{n} - i \gamma \Omega b_{n} - \omega^2(a_{n+1}-2 b_{n}+a_{n-1})&=&0\label{eq4}
\end{eqnarray}
where, without loss of generality, $a_{n}(b_{n})$ is the charge contained in the even (odd) capacitor, and
$\gamma=R/L$ and $\omega^2=1/L C$. Next, we employ a plane-wave ansatz: $a_{n}=A\ \exp(i k n)$, $b_{n}=B\ \exp(i k n )$. This leads to two linear, coupled equations:
\begin{eqnarray}
(\Omega^2-2 \omega^2-i \gamma \Omega) A + 2 \omega^2 \cos(k)\ B&=&0\nonumber\\
(\Omega^2-2 \omega^2+i \gamma \Omega) B + 2 \omega^2 \cos(k)\ A&=&0.
\end{eqnarray}
After imposing the vanishing of the determinant, we obtain $\Omega$:
\be
\Omega_{k}^{\pm}={1\over{\sqrt{2}}}\left( -\gamma^2+4 \omega^2 \pm \sqrt{(\gamma^2-4 \omega^2)^2-16 \omega^4\sin(k)^2}\right)^{1/2}
\ee
Simple analysis shows that the spectrum always contains complex eigenvalues for any gain/loss parameter value $\gamma$. This is in agreement with the results of Lazo et al. in the limit of an infinite latice(ref\cite{lazo}). This intrinsic instability seems to be generic to 1D systems.

Let us now introduce fractionality into our system. We notice that Eq.(\ref{eq4}) can be written as
\begin{eqnarray}
(-\Omega^2+ i \gamma \Omega) a_{n}-\omega^2(2 b_{n}-2 a_{n}+\Delta_{n} b_{n})&=&0\nonumber\\
(-\Omega^2- i \gamma \Omega) b_{n}-\omega^2(2 a_{n}-2 b_{n}+\Delta_{n} a_{n})&=&0,\label{eq7}
\end{eqnarray}
where $\Delta_{n}$ is the discrete Laplacian: $(\Delta_{n}) a_{n}= a_{n+1}-2 a_{n}+a_{n-1}$ and $(\Delta_{n}) b_{n}= b_{n+1}-2 b_{n}+b_{n-1}$. The transition to the fractional system is achieved by replacing each discrete Laplacian by its fractional version\cite{roncal} $(\Delta_{n})\rightarrow (\Delta_{n})^{\alpha}$, where 
\be
(-\Delta_{n})^\alpha C_{n}= \sum_{m\neq n} K^\alpha
(n-m) (C_{n}-C_{m}), \hspace{0.5cm} 0<\alpha<1\label{5}
\ee
where,
\be K^{\alpha}(m)={{4^\alpha \Gamma (\alpha+(1/2))\over{\sqrt{\pi}|\Gamma(-\alpha)|}}}{\Gamma(|m|-\alpha)\over{\Gamma(|m|+1+\alpha)}},\label{6}
\ee
and $\Gamma(x)$ is the Gamma function and $\alpha$ is the fractional exponent. Equations (\ref{eq7}) become
\begin{eqnarray}
(-\Omega^2+ i \gamma \Omega+2 \omega^2) a_{n}& &-\omega^2(2 b_{n}+\sum_{m}K^{\alpha}(n-m)(b_{m}-b_{n})) = 0\nonumber\\
(-\Omega^2- i \gamma \Omega+2 \omega^2) b_{n}& & -\omega^2(2 a_{n}+\sum_{m}K^{\alpha}(n-m)(a_{m}-a_{n}))
=0\nonumber
\end{eqnarray}
Next, as before, we insert the plane-wave ansatz $a_{n}=A\ \exp(i k n)$, $b_{n}=B\ \exp(i k n )$ arriving at a $2\times 2$ linear system for $A, B$:
\begin{eqnarray}
(-\Omega^2+ i \gamma \Omega+2\omega^2) A & &-(2\omega^2-4\omega^2\sum_{m=1}^{\infty}K^{\alpha}(m)\sin^2(k m/2)) B=0\nonumber\\
(-\Omega^2 - i \gamma \Omega+2\omega^2) B & &-(2\omega^2-4\omega^2\sum_{m=1}^{\infty}K^{\alpha}(m)\sin^2(k m/2)) A=0.\nonumber
\end{eqnarray}
Vanishing of the determinant of the system leads to the dispersion relation
\be
\Omega_{k}^{\pm}={1\over{\sqrt{2}}}\left( -\gamma^2 +4 \omega^2\pm \sqrt{\gamma^4 + 4 H(k)^2-8 \gamma^2\omega^2}  \ \right)^{1/2}\label{om}
\ee
where $H(k) = 2 \omega^2-4 \omega^2 \sum_{m=1}^{\infty} K^{\alpha}(m) \sin^2(k m /2)$.

\begin{figure}[H]
\includegraphics[scale=0.21]{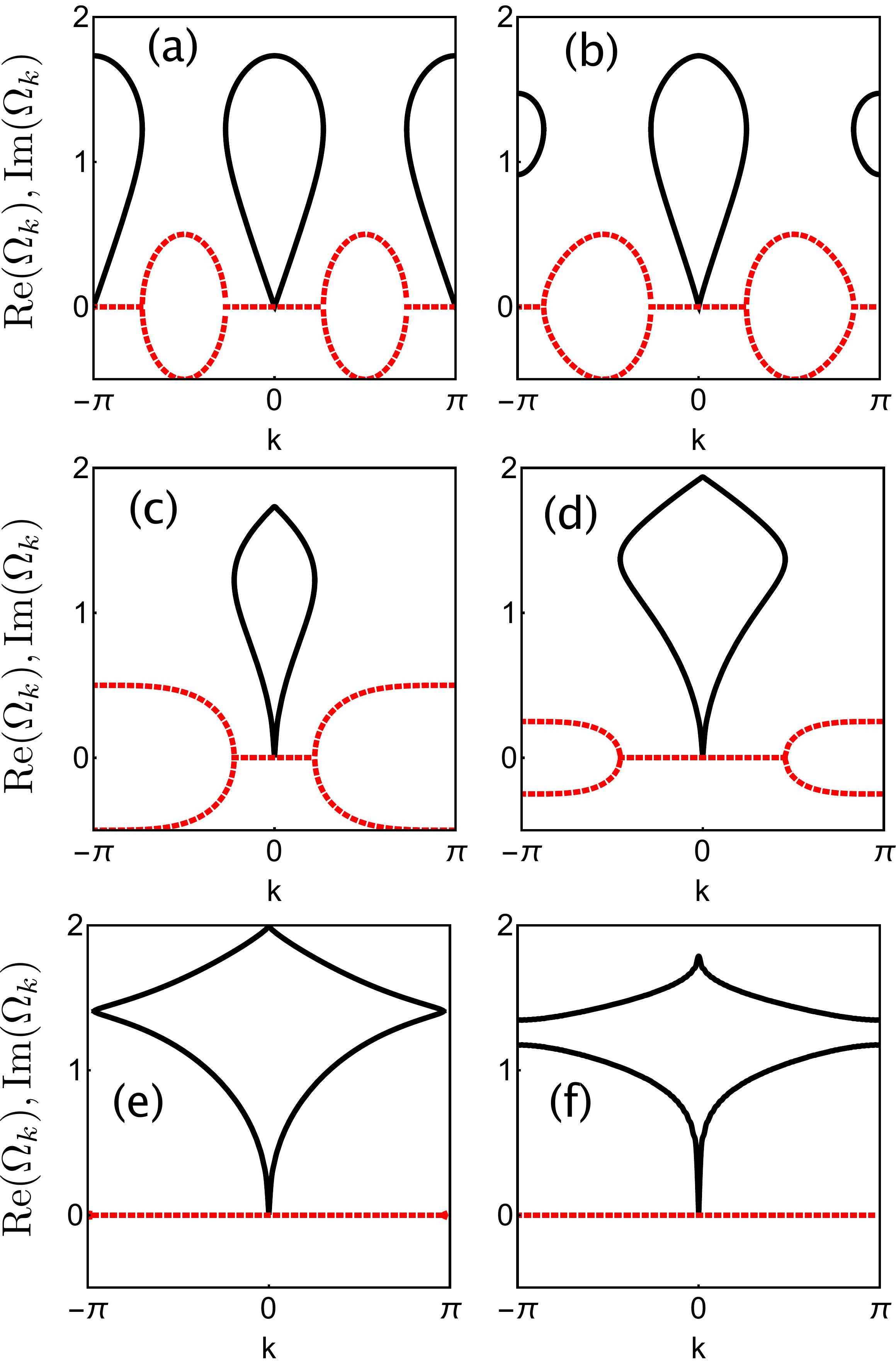}
 \caption{Real part (solid line) and imaginary part (dashed line) of the energy $\Omega_{k}$, as a function of wavevector $k$, for several gain/loss and fractional exponents: (a) $\gamma=1, \alpha=1$ (b) $\gamma=1, \alpha=0.9$ (c) $\gamma=1, \alpha=0.5$ (d) $\gamma=0.5, \alpha=0.5$ (e) $\gamma=0.2, \alpha=0.4$ (f) $\gamma=0.9, \alpha=0.1$ ($\omega=1$). In these  examples, only in cases (e) and (f) the spectrum is purely real leading to a stable dynamics. }
  \label{newfig2}
\end{figure}

Figure \ref{newfig2} shows some $\Omega_{k}$ plots for several $\gamma, \alpha$ values where we can
appreciate different regimes. We see that, in general, $\Omega_{k}$ is complex with real and imaginary parts of $\Omega_{k}$ alternating as wavevector values are swept. This is the unstable phase.
At small values of $\gamma$ and $\alpha$ the spectrum is real, and a gap is opened (Fig.2(f)). In this case, we are in the stable ${\cal PT}$ phase where  the dynamics is bounded. As the  fractional exponent is decreased further, the  real spectrum becomes flatter and flatter.

\begin{figure}[H]
\includegraphics[scale=0.21]{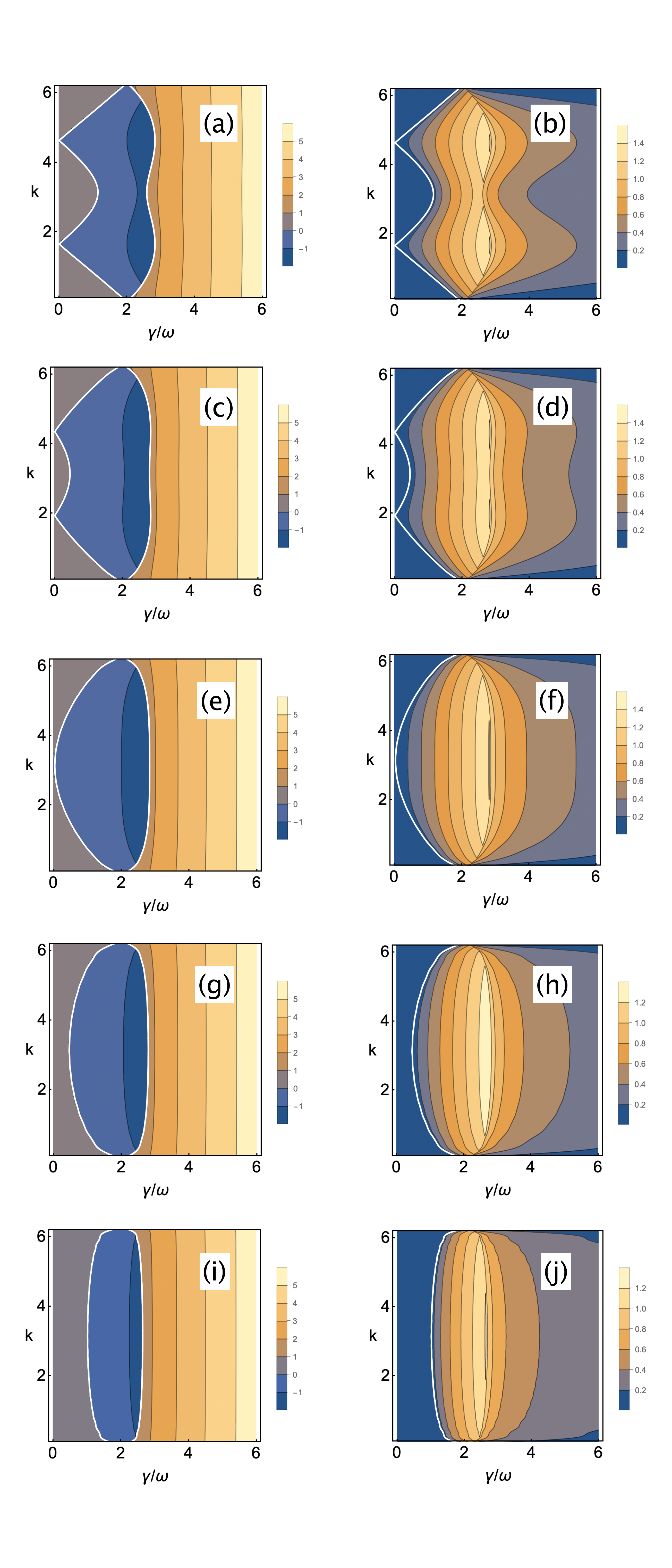}
 \caption{Im[$\Omega^{-}$] (left column) and Im[$\Omega^{+}$] (right column), as a function of the gain/loss parameter and wavevector, for several fractional exponents: $\alpha=0.9$ (a,b), $\alpha=0.8$ (c,d), $\alpha=0.5$ (e,f), $\alpha=0.2$ (g,h), $\alpha=0.1$ (i,j). Along the white line the imaginary part of the dispersion vanishes. Between the vertical axis $\gamma=0$ and the white line, the imaginary part of $\Omega^{\pm}$ is zero. ($\omega=1$). }
  \label{fig2}
\end{figure}

In figure \ref{fig2} we show a different, complementary view of the spectrum by displaying contours plots of $Im[\Omega^{\pm}(k)]$ in wavevector-and gain/loss parameter space for several fractional exponents $\alpha$. Stability occurs when $Im[\Omega^{\pm}(k)]$ is zero for both branches of the dispersion relation. As we can see, for fractional exponent greater than, approximately $1/2$, the imaginary part of $\Omega$ is non-zero, implying that our system is in the broken symmetry phase. This includes the standard case $\alpha=1$.  However, as $\alpha$ decreases further, a window of stability opens where $Im[\Omega^{\pm}(k)]=0$.

Using Eq.(\ref{om}) and $K^{\alpha}(m) \sim \alpha/m$ ($\alpha\rightarrow 0$), we can prove that
for a small fractional exponent, the stable region converges to $\gamma\leq 2 \omega$. Also, in this limit, we obtain a flat band spectrum with $\Omega_{k}=0$ and $\Omega_{k}=\sqrt{-\gamma^2+4 \omega^2}$.

The density of states $\delta(\Omega)=(1/N) \sum_k \delta(\Omega-\Omega_k)$ is complex in general since the spectrum is complex. We define partial densities of states for the real and imaginary part of the spectrum:
\begin{eqnarray}
\delta_{R}(\Omega) &=& (1/N) \sum_{\bf k} \delta(\Omega-Re[\Omega_k])\nonumber\\
\delta_{I}(\Omega) &=& (1/N) \sum_{\bf k} \delta(\Omega-Im[\Omega_k]).\label{16}
\end{eqnarray}

\begin{figure}[H]
\centering
\includegraphics[scale=0.21]{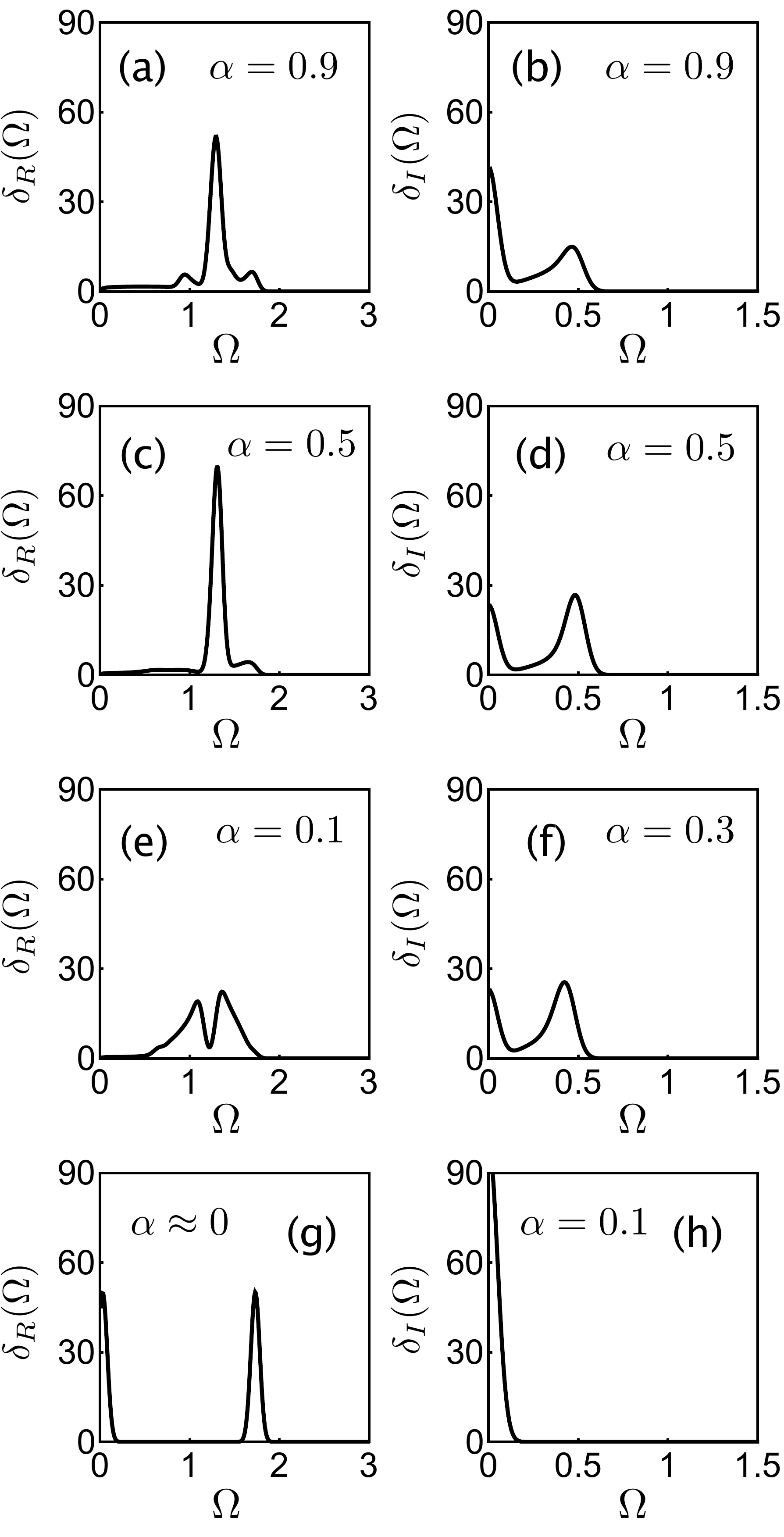}
 \caption{Left(right) column: Density of states of the real (imaginary) part of the spectrum,
 for $\gamma=1, \omega=1$. (a) and (b) $\alpha=0.9$; (c) and (d) $\alpha=0.5$; (e) $\alpha=0.1$;(f) $\alpha=0.3$; (g) $\alpha\approx 0$; (h) $\alpha=0.1$. }
  \label{fig4}
\end{figure}

By inserting the analytical expression (\ref{om}) into (\ref{16}) we compute numerically $\delta_{R}(\Omega)$ and $\delta_{I}(\Omega)$ for fixed values $\gamma=1=\omega$, and several fractional exponents $\alpha$, ranging from $\alpha\approx 1$ (non-fractional case) down to $\alpha\approx 0$. Results are shown in Fig.\ref{fig4}. As we can see, at small values of the fractional exponent ($\alpha\approx 0.1035$), the spectrum splits into two real bands which become completely flat in the limit $\alpha\rightarrow 0$, located at $\Omega=0$ and $\Omega=\sqrt{-\gamma^2+4 \omega^2}=\sqrt{-1^2+4\times 1^2}=1.73$. 
\begin{figure}[H]
\centering
\includegraphics[scale=0.21]{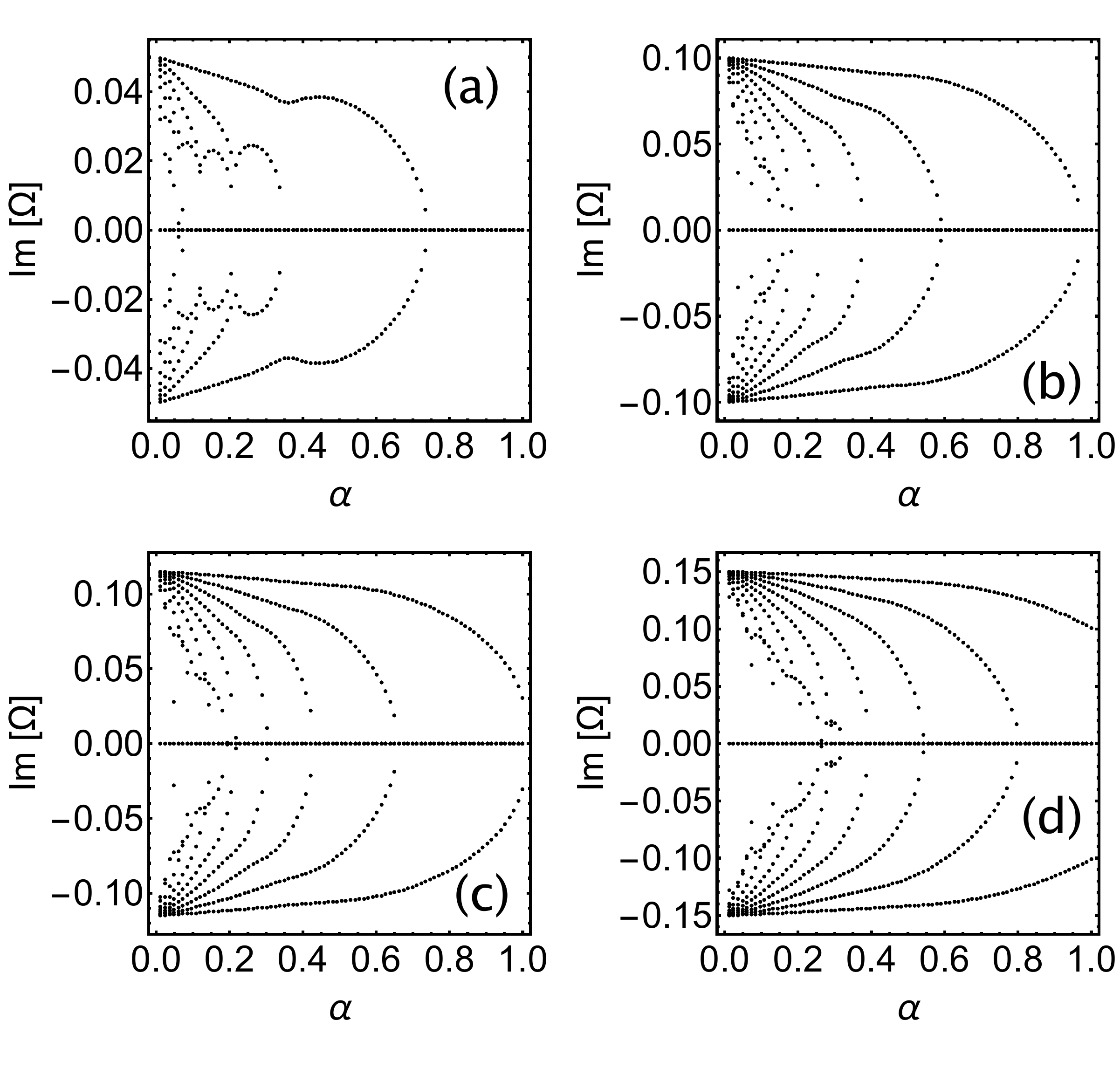}
 \caption{Imaginary part of the spectrum for a finite chain ($N=20$) vs the fractional exponent, for several gain/loss values: (a) $\gamma=0.1$, (b) $\gamma=0.2$, (c) $\gamma=0.23$, (d) $\gamma=0.3.$}
  \label{fig5}
\end{figure}
\begin{figure}[H]
\centering
\includegraphics[scale=0.21]{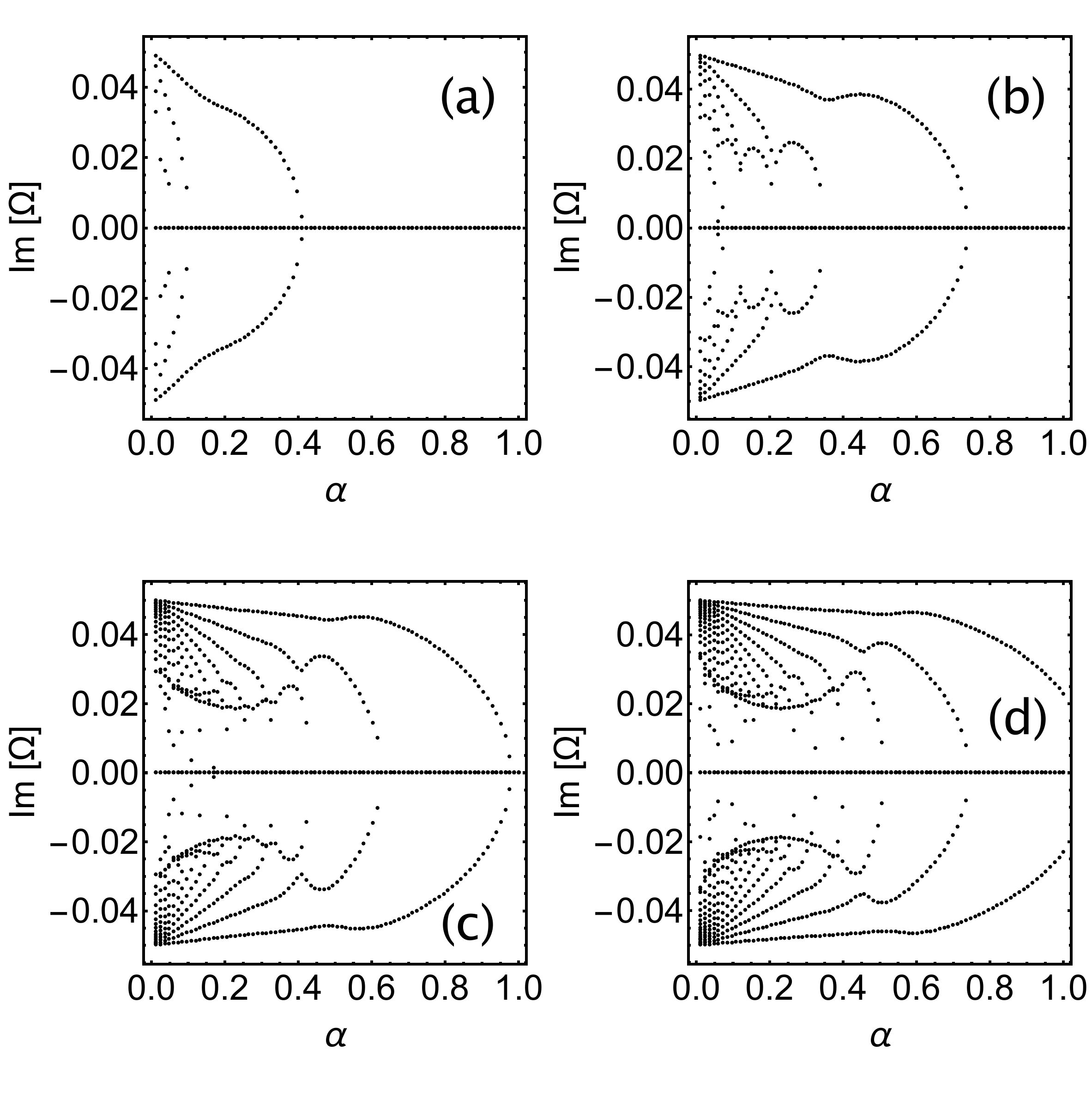}
 \caption{Imaginary part of the spectrum vs the fractional exponent, for several chain sizes: (a) $N=10$, (b) $N=20$, (c) $N=40$, (d) $N=50$\ ($\gamma=0.1$)}
  \label{fig6}
\end{figure}

All of the above treatment applies to an infinite lattice, but it is also instructive to look at what happens to a finite electrical transmission line. The starting point is the stationary equation
\be
-\Omega^2 Q_{n} + i \gamma \Omega Q_{n}-\omega^2 \sum_{m} K^{\alpha}(n-m) ( Q_{m} - Q_{n}) = 0
\ee
The eigenvalues $\{\Omega\}$ are extracted by using the augmented matrix method\cite{matrix}. Results are shown in Figs. 5 and 6, where we show scatter plots of the imaginary parts of the eigenvalues as a function of the fractional exponent, for given chain size $N$ and gain/loss parameter values $\gamma$. The case of a fixed chain size ($N=20$) and several $\gamma$ values is shown in Fig. 5. We see that, as $\gamma$ is increased the window where the eigenvalues are purely real shrinks. That is, an increase in gain/loss tends to destabilize the system. Also we notice that the stable region lies at relatively large $\alpha$ values, close to $\alpha=1$.

The complementary case of a fixed gain/loss parameter, and varying $N$ is shown in Fig.6. Here 
we notice that the stable region shrinks with increasing chain size, with the system becoming completely unstable around $N=40$. This is in agreement with previous results for the non-fractional ($\alpha=1$) DNLS equation for short ${\cal PT}$ chains\cite{short chains}, and can be traced back to the ability of a gain site to transfer its energy efficiently to a neighboring loss site. In our case, this leads to the estimate $N^{1-2 s}$ for the width of the stable window. The proof is not hard: At a gain site, the amplitude of the waves grows as $\exp(\beta t)$. This means that for a wave with wavevector $k$ to be stable, its group velocity needs $v_{k}$ to be greater than the speed at which the gain site accumulates energy that is, $v_{k}>\beta$, where $v_{k}=d \Omega_{k}/d k$. This must hold for all wavevectors. In particular, it must hold for the slowest mode, $k_{l}\ll 1$. For a periodic array $k_{l} = \pi/(N+1)$. Expansion of $v_{k}$ for small $k$ leads to
\be
v_{k}\approx\left\{
\begin{array}{rr}
(\omega/2) k \sum_{m}^{N} m^2 K^{\alpha}(m)\hspace{0.5cm}\mbox{upper band}\\
\\
\omega\   (\sum_{m}^{N} m^2 K^{\alpha}(m))^{1/2}\hspace{0.5cm}\mbox{lower band}\\
\end{array}
\right.\label{13}
\ee
The term $\sum_{m=1}^{N} m^2 K^{\alpha}(m)$ can be expressed in closed form as
\be
\sum_{m=1}^{N} m^2 K^{\alpha}(m)=
{(1+N) \alpha \Gamma(1+N-\alpha) \Gamma(2 \alpha) \sin(\pi \alpha)\over{\pi (1-\alpha) \Gamma(1+N+\alpha)}}.\label{14}
\ee
After replacement of Eq.(\ref{14}) into (\ref{13}), and using $k=k_{l}\approx \pi/N$, one obtains 
\be
\beta < {\alpha \Gamma(2 \alpha) \sin{\pi \alpha}\over{1-\alpha}}\ N^{1-2 \alpha}.
\ee
We see that the width of the stable region decreases more slowly for fractional exponents away from unity, the non-fractional limit.

{\em Discussion.}\ \   
We have examined the stability properties of a 1D chain of coupled electrical units, in the presence of ${\cal PT}$ symmetry and fractionality. In the absence of fractionality, previous work\cite{lazo} showed that this system is always in the broken symmetry phase. In this work,  we bring here a new element into play: the presence of fractional effects where the usual discrete Laplacian is replaced by its fractional form. This introduces nonlocal effects and long-range couplings. As we observed, the presence of fractionality opens a stability window in parameter space, where the eigenvalues are real. The behavior of this window seems to depend on whether the system is infinite or finite. For the infinite chain, the stability window lies at small gain/loss parameters and fractional exponents smaller than a critical value. On the contrary, for the finite case, the stability window lies at a small gain/loss and fractional exponents larger than a critical value. For fixed gain/loss and fractional exponent, the stability window shrinks with increasing system size. This last feature was observed previously for a 1D finite fractional magnetic metamaterial model\cite{short chains}. A rough explanation for the different behaviors could be as follows: The relevant sum $\sum_{m=1}^{\max} K^{\alpha}[m]$ has different limits for $\mbox{max}= \mbox{finite}$ and $\mbox{max}=\infty$, for $\alpha\rightarrow 0$:  $\sum_{m=1}^{\infty}K^{0}[m]\rightarrow 1/2$, while $\sum_{m=1}^{M} K^{0}[m]\rightarrow 0$, for finite $M$. Thus, we might expect different behaviors at small exponents. Also, we must not forget that  the range of the coupling is very long with $K^{\alpha}(m)\sim 1/|m|$ at small exponent and large distances. This means that for the finite case the system always sees' the boundaries of the chain. This perturbation could be responsible for the obliteration of a stability window at small exponents. Of course,  this disturbance effect is absent for the infinite chain.

The results for our classical, infinite lattice are qualitatively similar to the ones found previously for a quantum tight-binding chain\cite{molina}, suggesting some sort of universality of the behavior 
of the mutual influence of fractionality and ${\cal PT}$-symmetry in periodic discrete systems.

\vspace{0.1cm}

Given that ${\cal PT}$ management is of potential technological importance, these results encourage the experimental search of fractional effects in classical systems, such as electrical transmission lines or more generally, systems of coupled oscillators in higher dimensions. This depends on finding expressions for the fractional discrete Laplacians for the appropriate geometries, a nontrivial task under current investigation by the mathematical community.

\acknowledgments
This work was supported by Fondecyt Grant 1200120.


\begin{thebibliography}{99}

\bibitem{bender1}
C. M. Bender and S. Boettcher, Real Spectra in Non-Hermitian Hamiltonians Having ${\cal P\cal T}$ Symmetry, Phys. Rev. Lett. {\bf 80}, 5243 (1998).

\bibitem{bender2}
C. M. Bender, D. C. Brody, and H. F. Jones, Complex Extension of Quantum Mechanics, Phys. Rev. Lett. {\bf 89}, 270401 (2002).

\bibitem{optics1}
R. El-Ganainy, K. G. Makris, D. N. Christodoulides, and Z. H. Musslimani, Theory of coupled optical ${\cal P}{\cal T}$-symmetric structures, Opt. Lett. {\bf 32}, 2632 (2007).

\bibitem{solid1}
N. Hatano and D. R. Nelson, Localization Transitions in Non-Hermitian Quantum Mechanics, Phys. Rev. Lett. {\bf 77}, 570 (1996).

\bibitem{solid2}
 Y. N. Joglekar, D. Scott, M. Babbey, and Avadh Saxena, Robust and fragile 
${\cal P}{\cal T}$-symmetric phases in a tight-binding chain
Phys. Rev. A {\bf 82}, 030103 (2010).

\bibitem{optics2}
Z. H. Musslimani, K. G. Makris, R. El-Ganainy, and D. N. Christodoulides, Optical Solitons in 
${\cal P}{\cal T}$ Periodic Potentials, Phys. Rev. Lett. {\bf 100}, 030402 (2008).

\bibitem{optics3}
K. G. Makris, R. El-Ganainy, D. N. Christodoulides, and
Z. H. Musslimani, Beam Dynamics in ${\cal P}{\cal T}$ Symmetric Optical Lattices, Phys. Rev. Lett. {\bf 100}, 103904 (2008).

\bibitem{optics4}
A. Guo, G. J. Salamo, D. Duchesne, R. Morandotti, M. Volatier-Ravat, V. Aimez, G. A. Siviloglou, and D. N.
Christodoulides, Observation of ${\cal P}{\cal T}$-Symmetry Breaking in Complex Optical Potentials, Phys. Rev. Lett. {\bf 103}, 093902 (2009).


\bibitem{optics5}
C. E. R\"{u}ter, K. G. Makris, R. El-Ganainy, D. N. Christodoulides, M. Segev, and D. Kip, Observation of parity-time symmetry in optics, Nat. Phys. {\bf 6}, 192 (2010).

\bibitem{circuits}
J. Schindler, Ang Li, M. C. Zheng, F. M. Ellis, and T.
Kottos, Experimental study of active LRC circuits with 
${\cal P}{\cal T}$ symmetries, Phys. Rev. A {\bf 84}, 040101 (2011).

\bibitem{magnetics1}
N. Lazarides and G. P. Tsironis, 
Gain-Driven Discrete Breathers in ${\cal P}{\cal T}$ Symmetric Nonlinear Metamaterials,
Phys. Rev. Lett. {\bf 110}, 053901 (2013).

\bibitem{magnetics2}
Mario I. Molina, Bounded dynamics in finite ${\cal PT}$-symmetric nonlinear metamaterials,
Phys. Rev. A {\bf 89}, 033201 (2014).

\bibitem{lazo}
E. Lazo and F. Humire, ${\cal PT}$ symmetric direct electrical transmission lines: Localization behavior, Phys. Rev. E {\bf 100}, 022221 (2019).

\bibitem{experiment2}
A. Guo et al., Observation of ${\cal P}{\cal T}$-Symmetry Breaking in Complex Optical Potentials, Phys. Rev. Lett. {\bf 103}, 093902 (2009).

\bibitem{experiment3}
A. Szameit, M. C. Rechtsman, O. Bahat-Treidel, and M.
Segev, ${\cal P}{\cal T}$-symmetry in honeycomb photonic lattices, Phys. Rev. A {\bf 84}, 021806 (2011).

\bibitem{quantum}
L. A. Caffarelli, A. Vasseur, Drift diffusion equations with fractional diffusion and the quasi-geostrophic equation, Ann. Math. {\bf 171}, 1903 (2010).

\bibitem{kinetics1}
R. Metzler, J. Klafter, The random walk's guide to anomalous diffusion: a fractional dynamics approach, Phys. Rep. {\bf 339}, 1-77 (2000).

\bibitem{kinetics2}
I. M. Sokolov, J. Klafter, A. Blumen, Fractional kinetics, Physics Today {\bf 55}, 48 (2002).

\bibitem{kinetics3}
G. M. Zaslavsky, Chaos, fractional kinetics, and anomalous transport, Phys. Rep. {\bf 371}, 461 (2002).


\bibitem{strange}
M. F. Shlesinger, G. M. Zaslavsky and J. Klafter, Strange kinetics, Nature {\bf 363}, 31 (1993).

\bibitem{frac1}
N. Laskin, Fractional quantum mechanics, Phys. Rev. E {\bf 62}, 3135 (2000).

\bibitem{frac2}
N. Laskin, Fractional Sch\"{o}dinger equation, Phys. Rev. E {\bf 66}, 056108 (2002).

\bibitem{levy}
N. C. Petroni and M. Pusterla, Levy processes and Schrodinger equation, Physica A {\bf 388}, 824 (2009).

\bibitem{plasmas}
M. Allen, A fractional free boundary problem related to a
plasma problem, Commun. Anal. Geom. {\bf 27}, 1665 (2019).

\bibitem{cardiac}
A. Bueno-Orovio, D. Kay, V. Grau, B. Rodriguez and K.
Burrage, Fractional diffusion models of cardiac electrical propagation: role of structural heterogeneity in dispersion of repolarization, J. R. Soc. Interface {\bf 11}, 20140352 (2014).

\bibitem{invasions}
J. Manimaran, L. Shangerganesh, Amar Debbouche, and Valery Antonov, Numerical Solutions for Time-Fractional Cancer Invasion System With Nonlocal Diffusion,  Frontiers in Physics {\bf 7},
1-16 (2019).

\bibitem{epidemics}
 S. Pooseh, H. S. Rodriguez and Delfim F. M. Torres, Fractional Derivatives in Dengue Epidemics, AIP Conference Proc. {\bf 1389}, 739 (2011).

\bibitem{molina}
Mario I. Molina, Interplay of fractionality and ${\cal PT}$-symmetry on a 1D lattice,
Phys. Lett. A {\bf 449}, 128336 (2022).

\bibitem{roncal}
Oscar Ciaurri, Luz Roncal, Pablo Raul Stinga, Jose L. Torrea, Juan Luis Varona,
Nonlocal discrete diffusion equations and the fractional discrete Laplacian, regularity and applications, Adv. Math. {\bf 330}, 688 (2018).




\bibitem{matrix}
Matrices applied to the motions of damped systems, 
W. J. Duncan and A. R. Collar, Phil. Mag. {\bf 19}, 197 (1935).

\bibitem{short chains}
M. I. Molina, Bounded dynamics in finite ${\cal PT}$-symetric magnetic metamaterials, 
Phys. Rev. E {\bf 89}, 033201 (2014). 







\end{thebibliography}
\end{document}